\documentclass[a4paper,12pt]{article}

\usepackage[utf8]{inputenc}
\usepackage[english]{babel}

\usepackage{graphicx}
\usepackage[table,xcdraw]{xcolor}
\usepackage{hyperref}
\usepackage[english]{babel}
\usepackage{amsmath}
\usepackage[a4paper, top=2.5cm, left= 2.5cm, right=2.5cm, bottom=2cm]{geometry}
\usepackage{csquotes}

\usepackage[gen]{eurosym}

\usepackage[round,sectionbib,sort]{natbib}
\setcitestyle{aysep={}}

\usepackage[T1]{fontenc}
\usepackage[hang]{footmisc}
\usepackage{lmodern}

\usepackage{listings}

\usepackage{color}

\definecolor{codegreen}{rgb}{0,0.6,0}
\definecolor{codegray}{rgb}{0.5,0.5,0.5}
\definecolor{codepurple}{rgb}{0.58,0,0.82}
\definecolor{backcolour}{rgb}{0.95,0.95,0.92}

\lstdefinestyle{mystyle}{
  backgroundcolor=\color{backcolour},   commentstyle=\color{codegreen},
  keywordstyle=\color{magenta},
  numberstyle=\tiny\color{codegray},
  stringstyle=\color{codepurple},
  basicstyle=\footnotesize,
  breakatwhitespace=false,         
  breaklines=true,                 
  captionpos=b,                    
  keepspaces=true,                 
  numbers=left,                    
  numbersep=5pt,                  
  showspaces=false,                
  showstringspaces=false,
  showtabs=false,                  
  tabsize=2
}

\begin{document}

\thispagestyle {empty}
\begin{center}

\quad

\vspace{0cm}

\vspace{0.3cm}
\textbf{Implementation of a cost-benefit analysis}\\
\textbf{of Demand-Responsive Transport with a}\\
\textbf{Multi-Agent Transport Simulation}\\

\vspace{0.3cm}

Conny Grunicke$^{\text{a},\text{b}}$ \\
Jan Christian Schlüter$^{\text{b,c,*}}$ \\
Jani-Pekka Jokinen$^{\text{b}}$ \\

\vspace{0.3cm}

$^{\text{a}}$ Chair of Statistics, Department of Economics, Georg-August-University of Göttingen, Humboldtallee 3, 37073 Göttingen, Germany\\

\vspace{0.2cm}

$^{\text{b}}$ Department of Dynamics of Complex Fluids, Max-Planck-Institute for Dynamics and Self-Organization,   Am Fassberg 17, 37077 Göttingen,
Germany\\

\vspace{0.2cm}

$^{\text{c}}$ Institute for the Dynamics of Complex Systems, Faculty of Physics, Georg-August-University of Göttingen, Friedrich-Hund-Platz 1, 37077
Göttingen, Germany\\
\end{center}

\section*{Abstract}
In this paper, the technical requirements to perform a cost-benefit analysis of a Demand
Responsive Transport (DRT) service with the traffic simulation software MATSim are
elaborated in order to achieve the long-term goal of assessing the introduction of a DRT
service in Göttingen and the surrounding area. The aim was to determine if the software
is suitable for a cost-benefit analysis while providing a user manual for building a basic
simulation that can be extended with public transport and DRT. The main result is that
the software is suitable for a cost-benefit analysis of a DRT service. In particular, the
most important internal and external costs, such as usage costs of the various modes
of transport and emissions, can be integrated into the simulation scenarios. Thus, the
scenarios presented in this paper can be extended by data from a mobility study of
Göttingen and its surroundings in order to achieve the long-term goal. This paper is aimed at transport economists and researchers who are not familiar with MATSim, to provide them with a guide for the first steps in working with a traffic simulation soft-ware.

\ \\ \textit{Keywords:} \textbf{Transport Simulation; MATSim; Agent-Based Modeling; Demand Responsive Transport; Cost-Benefit Analysis}
\addcontentsline{toc}{section}{Abstract}

\vspace{0.1cm}

\ \\ {*}Corresponding author at: Max-Planck-Institute for Dynamics and Self-Organization, Department of Dynamics of Complex Fluids,  Am Fassberg 17, 37077 Göttingen, Germany 
\ \\ E-mail address: jan.schlueter@ds.mpg.de\\






\clearpage
\pagenumbering{arabic}
\setcounter {page}{1}


\section{Introduction}


Due to the significant increase in the volume of traffic in recent decades and the dominant position of motorized private transport, the negative effects on people and the environment have intensified \citep{UBAFahrleistung}. To reduce greenhouse gas emissions and prevent climate change, new solutions and innovations are needed in sectors such as transport. For this reason, sustainable transport systems, such as cycling, public transport, and new mobility concepts, are of increasing interest to scientists of different disciplines and public decision-makers \citep{gossling2015transport,johansson2017evaluation}.
The economic evaluation of transport measures and policies is used to assess their economic efficiency and benefits. To assess the overall welfare impact of such measures and policies, many countries frequently conduct a cost-benefit analysis as a guideline for decisions on infrastructure projects \citep{bristow2000transport,mackie2013international}.
To determine the overall impact on welfare, this assessment method examines the extent of the social and economic impact. An essential feature of the analysis is that it also monetarises negative external effects, such as congestion, air ,and noise emissions, and thus reflects not only private costs but also social costs \citep{damart2009uses}. In addition, social acceptance has a significant influence on the prospects of successful implementation of a measure or a new concept. Therefore, this implies that a consideration of the individual benefits of transport infrastructure is very helpful for a comprehensive assessment.
\ \\A paper by \citet{kickhofer2009methodik} argues that multi-agent transport simulation is able to illustrate those individual changes in the welfare level which results from changes in transportation mode, travel times or travel costs. The use of simulations, in general, is not uncommon in the transport sector, as these make it possible to predict the long-term effects of new mobility concepts, even before their introduction and the investments associated with them. The main difference between conventional transport simulations and the agent-based simulation is the representation of changes in utility on an individual level. However, the use of multi-agent transport simulation is relatively new and has not yet been used for a cost-benefit analysis of new mobility concepts. Part of these new mobility concepts are flexible intelligent transport systems. One realization of such a flexible service is Demand-Responsive Transport (DRT). DRT dynamically adapts to travel requirements by operating a fleet of vehicles without fixed routes and schedules. This allows journeys to be made without transferring, as either a door-to-door or a corner-to-corner service is offered. As there seems to be no study using agent-based simulation for a cost-benefit analysis of new flexible transport services such as DRT, the contribution of this paper is to elaborate the technical requirements for it. Therefore, the intention of this project paper is to prepare the long-term goal of simulating the traffic in Göttingen and evaluating the offer of a DRT service in Göttingen and the surrounding area. The traffic in Göttingen shall serve as an example. The current traffic situation represents the baseline scenario. In a second step, this will be compared with a scenario in which there is an additional offer of DRT in the Göttingen area. This comparison is intended to build the foundation for conducting a cost-benefit analysis and assesses the choice of means of transport and its impacts on travel times, internal and external costs such as congestion, noise, and air pollution. It will then be investigated whether negative external effects can be reduced by the offer of a DRT service as a traffic measure. 
\ \\Consequently, this paper has two aims. The main goal of this project paper is to elaborate the technical requirements, i.e. to get to know the multi-agent transport simulation software MATSim and to perform first simulations. Since this software is Java-based, the development of Java knowledge is also a technical requirement. Secondly, further steps for the implementation of a simulation with real traffic data and the necessary extensions to perform a cost-benefit analysis are discussed. As a first approximation, only a certain route is simulated and not the whole city area. The exemplary route is from the city centre to the Fassberg in Göttingen, as this is the commuting route for many employees of the Max Planck Institute. This first simulated route represents the preparation for a traffic simulation for the entire city area.
\ \\In the following section, a brief overview of MATSim with the specific advantages and disadvantages of its application is given. This is followed by a description of the simulation model. Thereby, the data collection and the technical aspects for the generation of all relevant information for the simulation of the baseline situation and the DRT scenario are considered. Section 4 describes the results of the simulations, particularly the resulting reaction patterns. Finally, the next steps for further work are presented and the conclusion is drawn. 

\section{Multi-Agent Transport Simulation}

This paper uses the open-source agent-based microsimulation MATSim. In traffic-related areas, the use of agent-based simulation is becoming increasingly popular, and several examples of agent-based approaches can be found in the literature in recent years \citep{bischoff2016simulation, kaddoura2015optimal}. 
This section provides a general overview of MATSim (section \ref{sec:What is MATSim?}), including the specific advantages and disadvantages of its application (section \ref{sec:Why using MATSim?}). Thereby the chapter tries to be as general as possible at the beginning. In the following chapters, relevant features and concrete examples will be discussed in more detail.

\subsection{What is MATSim?}\label{sec:What is MATSim?}

\begin{figure}
\includegraphics[width=0.9\linewidth]{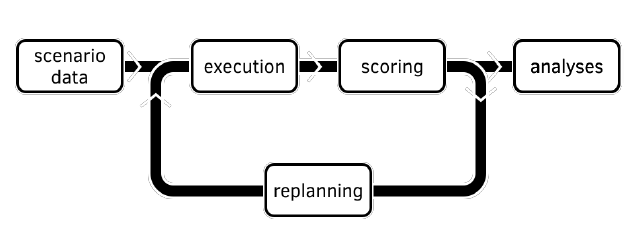}
\caption{The building blocks of MATSim \citep{horni2016multi}}
\label{figure:1}
\end{figure}

The traffic simulation MATSim is a Java-based open-source project jointly developed by the Technical University of Berlin and the ETH Zurich. The MATSim software allows a microscopic simulation of agent behaviour with high computing power as well as the possibility to investigate large traffic scenarios \citep{bischoff2016simulation}. In this context, the term microsimulation means that each person is simulated as an independent agent with any number of individual characteristics. MATSim offers a set of modules that can be adapted or extended according to individual needs: 
\begin{itemize}
    \item{Modelling of transport demand}
    \item{Agent-based microsimulation}
    \item{Iterative evaluation of the daily plans (scoring)}
    \item{Variation of daily plans (replanning)}
    \item{Methods for evaluating the results}
\end{itemize}
\ \\Together, these modules form an integrated optimization system referred to as the MATSim Circle. This optimization system is based on a co-evolutionary principle. Each agent repeatedly optimizes a daily activity plan, competing for space-time slots with all other agents. This co-evolutionary principle is implemented in MATSim as a loop. It begins with the modelling of an initial demand, which results from the daily activity chains of the persons in the study area. In MATSim these modeled persons are referred to as agents. The activity chains are usually derived from empirical data or discrete decision modelling. Agents' responsiveness to other agents competing for space-time slots enables them to individually optimize the initial demand during the iterations\footnote{The simultaneous simulation of all agents leads to interactions, thus longer travel times or even traffic jams can be modelled.}. Thereby each agent has a memory with a fixed number of daily plans, whereby each plan consists of a daily activity chain and a corresponding score. The score can be interpreted as an economic utility. During all iterations, every agent selects a plan from its memory. This selection depends on the scoring of the plans, which are calculated after each run based on the performance of the executed plans. A specified share of agents can copy the selected plan and change the duplicate, which is the replanning phase of the loop. Replanning is possible for different dimensions. There is the possibility to change the travel time, the route, the modes of transport and the destination. Besides, there are various replanning strategies for adapting plans, ranging from random mutations to approximate suggestions to best-response answers. Each iteration ends with the evaluation of the agents' experiences from the selected daily plans (scoring). The applied scoring function assigns a positive utility for the performance of activities and a negative utility for traveling, monetary costs such as tolls or fares and late arrival or early departure. However, the scoring function can be changed and thus customized to the individual needs of the user \citep{horni2016multi}. This repetitive process typically leads to an approximate stochastic user equilibrium \citep{nagel2012agent}. Finally, MATSim contains some methods for evaluating the results directly out of the box. 
However, most analyses are very specific and must be implemented according to the user's requirements. Figure \ref{figure:1} represents the loop described above with the configurable number of iterations that a MATSim run consists of.

\ \\The basic input data for a MATSim simulation are on the one hand a street graph\footnote{The underlying structure of a street graph can be modelled mathematically as a graph, thus graph theory is applied in the broadest sense.} that maps the relevant infrastructure of the study area and on the other hand the initial traffic demand already mentioned above. Beyond these minimum requirements, additional data can be added to the model to improve detail accuracy \citep{horni2016multi}. The model described in this paper also contains data exceeding the minimum requirements. Details of the additional data used are described in section \ref{sec:Simulation Model for G\"ottingen}. 

\subsection{Why using MATSim?}\label{sec:Why using MATSim?}

After the functionality of MATSim was presented above, the advantages and disadvantages of using this software can be summarized. The conceptual approach of MATSim differs fundamentally from conventional traffic simulation software. While traditional simulation software only assigns an aggregated demand to the respective traffic network and thus only static flows on the routes can be represented in predefined time intervals, the microscopic, behavior-based approach of MATSim enables a much lower spatial, individual and time-dependent analysis \citep{kickhofer2009methodik}. Thus, it can be concluded that through the individual simulation of each traveler, agent-based modeling can describe the demand side of the model in great detail. This is an advantage that traditional transport modeling cannot offer \citep{ciari2008concepts}. Based on the individual simulation there is another advantage of the MATSim software. Namely, the consideration of behavioural aspects, which can be relevant in the transport sector, e.g. the choice of route, travel time or mode of transport \citep{rieser2010adding}. Behavioural rules for agents are defined at the individual level, but the overall response to different scenarios is the result of both the individual response to such measures and the interaction between all simulated individuals \citep{ciari2008concepts}. Because of the representation of the agents with attributes and a certain logic it is a very intuitive possibility to simulate the traffic. Another factor is that the evaluation of the profitability of DRT providers requires a microsimulation. Aggregation-based methods are unsuitable because the average demand ignores the effects of the distribution in the temporal and/or spatial dimension. In addition, the agent paradigm can also be expanded to the supply side of the transport system, allowing an even more comprehensive representation of the overall system \citep{cich2017modeling}. Altogether, MATSim is able to realistically map the dynamic traffic volume due to the described advantages, provided that the necessary traffic data is available. Furthermore, it should be noted that MATSim is a fast dynamic simulation software with high computing power even for large traffic scenarios\citep{agarwal2019calibration}. MATSim also offers high adaptability due to its modular design. There are many plug-in points to extend functionality. The software can be extended by already written open source plugins\footnote{In these cases, there is often good documentation.} as well as by self-written algorithms. One drawback, however, is that there is no easy to use interface for the self-written customization, but very good Java knowledge is required to make all individual customization. The main disadvantage, however, is the implementation effort, which increases significantly with each new detail, i.e. with an increasing level of detail and accuracy of the mapping.

\section{Simulation Model for G\"ottingen} \label{sec:Simulation Model for G\"ottingen}

As already mentioned, this project paper is intended to prepare the long-term goal of traffic simulation and the evaluation of the offer of a DRT service in Göttingen and the surrounding area. For this reason, the road network of Göttingen, Bovenden, and Rosdorf has been selected as the data to be used for the forthcoming work will cover this region of interest. These data come from a household survey on mobility behaviour conducted by the City of Göttingen in 2015/2016 \citep{MobilitaetGoettingen}. This included the availability of means of transport, the purposes of the journeys made, the time distribution of journeys over the day and much more. All this information is valuable input data for the transport simulation and makes it possible to investigate whether the simulation carried out is representative and thus close to the actual traffic conditions in Göttingen. This section describes the simulation of the exemplary route from the city centre to the Fassberg in Göttingen which was chosen to get to know the technical requirements for a simulation with MATSim. Since only a small part of the population was simulated, there were no requirements to represent the population completely or representative for this project. In the following, the collection of the necessary input data for the traffic model is described. Subsequently, the baseline scenario, as well as the DRT scenario, are presented. The focus continues to be on the technical aspects.

\subsection{Data collection} \label{sec:Data collection}

\begin{figure}[h!]
\centering
\includegraphics[width=0.65\linewidth]{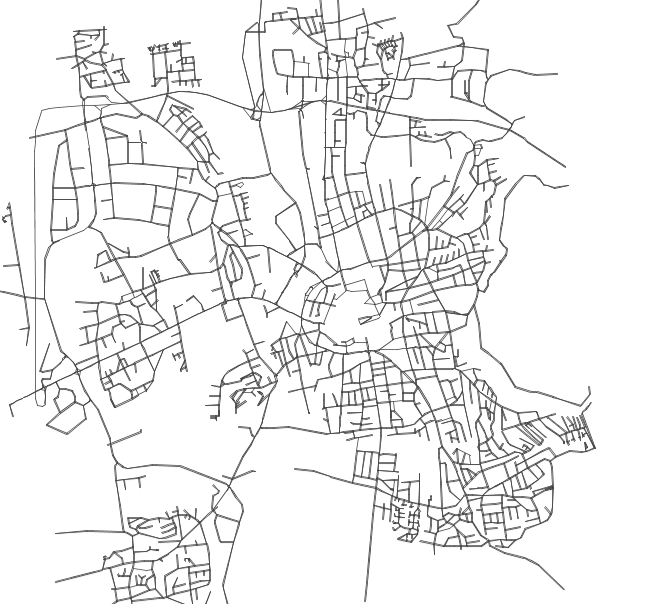}
\caption{Visualization of the G\"ottingen Network}
\label{figure:2}
\end{figure}

During the first phase of the project, the necessary database for the creation of an agent-based traffic model was developed. In general, the transport model consists of two components, transport supply, and transport demand. The transport supply is represented by the transport network and the transport infrastructure (e.g. public transport lines). Traffic demand is represented by mobility data or agent activity plans. OpenStreetMap (OSM) data served as the foundation for mapping the road network of Göttingen, Rosdorf and Bovenden. These data were supplemented by local knowledge of the road network with the program JOSM - Java OpenStreetMap Editor. For OSM data to be suitable for routing by MATSim, it must be converted to a xml data format. MATSim provides the Java class \textit{OsmNetworkReader} for this purpose. This reader takes the OSM data and generates a MATSim network according to predefined settings. In addition, the simulation software uses a \textit{NetworkCleaner} and a \textit{NetworkSimplifier} to make the network usable for a simulation and to reduce the computational effort. When creating the MATSim road network, Gauß-Krüger coordinates were used because the MATSim community recommends using a regional coordinate system and Gauß-Krüger is mainly used in Germany \citep{horni2016multi}. After this conversion, the network contains 7863 road links and 3680 nodes. A representation of the network is illustrated in figure \ref{figure:2}. 
\ \\In order to map public transport, GTFS\footnote{General Transit Feed Specification is a widely used format for public transport schedules and geographical information.} data was necessary. These data were provided by a subcontractor of the transport authority in Göttingen. The collected data were used by the MATSim package \textit{PT2MATSim}. The exact description of this process is explained in section \ref{sec:Baseline Situation}. The data on transport demand will, in the long run, be based on a household survey on mobility behaviour in Göttingen carried out by the City of Göttingen in 2015/2016, which was made available for the project. However, since only one exemplary route is simulated for this project, the transport demand was generated randomly, so that the population modelling is initially only approximate. Thus, there was no requirement for completeness or representativity. The modelling of transport demand is discussed in greater detail in section \ref{sec:Baseline Situation}. 


\newpage
\subsection{Baseline Situation} \label{sec:Baseline Situation}

A crucial step in assessing the benefits and costs of a transport project is to define the baseline situation. In the long term, the current traffic situation will represent the baseline scenario. However, for this paper, the base case is the commuting scenario from the city centre to the Fassberg where only the car and the public transport is allowed. As already mentioned in section \ref{sec:What is MATSim?}, there are some basic input data for MATSim: the street graph and the initial transport demand. It is already explained in the previous section how to configure the road network for simulation. This transport network data contains information about the capacities, maximum speeds, link length and permitted means of transport on the individual routes. In order to map the base case of road traffic in Göttingen correctly, it was necessary to create a multimodal road network on which public transport can also run. Apart from the multimodal road network, some other input data beyond the basic input data is necessary to simulate the public transport in MATSim. Essentially, there are two relevant input files. The first is the \textit{TransitSchedule}, which contains all public transport lines, routes, stops and departures. It describes all planned trips of public transport vehicles in the system. Thereby, each vehicle is simulated separately in the traffic flow simulation. Time delays may occur for transit vehicles, and actual departures may differ from the planned schedule, depending on the type of vehicle and the number of passengers boarding and alighting. The second file is the \textit{TransitVehicles}, which contains all vehicle types and all vehicles actually used. As mentioned above, the \textit{PT2MATSim} package developed by \citet{poletti2016public} was used to generate these necessary input data for public transport. It is based on an algorithm that makes it possible to see the effects of the interaction of buses and motorized individual traffic in a traffic simulation. The implementation provides tools to convert GTFS data into MATSim transit schedules and to create a multimodal MATSim network with OSM. Therefore, as a first step, the GTFS converter was used. 
The converter's main job is transforming GTFS trips to MATSim transit routes. For this purpose, the coordinates of the stops are extracted from the GTFS data. The converter is called by the Java main method \textit{Gtfs2TransitSchedule}, followed by the folder where the GTFS files are located. Afterwards, the date for which the routes are to be extracted is specified\footnote{MATSim normally simulates only one day, so a sample day was extracted from the GTFS feed.}. Then the output coordinate system\footnote{EPSG codes are a widely used means to clearly indicate the spatial reference of geodata. EPSG:31467 is the code for Gauß-Krüger coordinates.} and lastly the output files of the transit plan and the transit vehicles are specified. The code for the implementation of the GTFS converter is presented in listing \ref{lst:Gtfs}. 

  
\ \\\begin{lstlisting}[language=xml, basicstyle=\footnotesize\sffamily, numbers=left, escapechar=\%, numberstyle=\tiny, frame=tb, tabsize=4, lineskip=-0.1ex, caption={Gtfs2TransitSchedule},label=lst:Gtfs]
 java -cp pt2matsim-19.2-shaded.jar 
 org.matsim.pt2matsim.run.Gtfs2TransitSchedule
 "path to the GTFS data" 
 "date" "EPSG:code" 
 "path to output transit schedule" 
 "path to output transit vehicles"
\end{lstlisting}

\ \\For this paper, only the schedules for the bus lines 21, 22 and 23 in Göttingen were converted, because only these connect the inner city with the Fassberg and are therefore relevant for the exemplary route. However, those \textit{TransitSchedules} are 'unmapped' after using the GTFS converter, which means that the schedules do not contain information about the links used by the vehicles. They simply include the stop order for transit routes. The creation of these links (i.e. the route taken by a vehicle in a network) is called 'mapping', a process performed by the \textit{PublicTransitMapper}. It implements an algorithm that compiles a series of link candidates for each stop by selecting the nearest \textit{n} links from the coordinates of the stop. From these, the algorithm calculates the least cost path from the transit route’s first to its last stop\footnote{The least cost path is calculated from the path travel cost plus half the travel costs of the two link candidates the path connects. Then, the travel cost between the candidates (i.e. the travel cost on the link) is multiplied by four. This is done to avoid the same link pointing to two stops. The travel cost on a network link is normally length or travel time \citep{poletti2016public}.}. In the case that no link candidate can be found within the predefined distance an artificial loop connection is created since all stops need to be referenced to a link for the simulation to work \citep{poletti2016public}. This situation occurred in the base case of this project as the OSM data for the Göttingen inner city was not complete. This is probably due to the fact that the city centre is mostly car-free. The resulting artificial links, however, represent the centre's road network very well. The reason for this is that the density of the stops in this area is fairly high and the stops are relatively close to each other. To map a transit schedule to a MATSim network, a configuration file with all input parameters is necessary. The code for the creation of a default configuration file is shown in listing \ref{lst:defaultconfig}. 
\ \\\begin{lstlisting}[language=xml, basicstyle=\footnotesize\sffamily, numbers=left, escapechar=\%, numberstyle=\tiny, frame=tb, tabsize=4, lineskip=-0.1ex, caption={Creation of a default configuration file},label=lst:defaultconfig]
java -cp pt2matsim-19.2-shaded.jar
org.matsim.pt2matsim.run.CreateDefaultPTMapperConfig
"path to the default configuration file"
\end{lstlisting}
\newpage
\ \\In this configuration file, the \textit{TransitSchedule} file created by the converter as well as the network file must be integrated as input files. The output files also need to be specified. Afterwards, the schedule can be mapped to the network using the \textit{PublicTransitMapper}. This is done by executing the code from listing \ref{lst:PTMapper}.
\ \\\begin{lstlisting}[language=xml, basicstyle=\footnotesize\sffamily, numbers=left, escapechar=\%, numberstyle=\tiny, frame=tb, tabsize=4, lineskip=-0.1ex, caption={PublicTransitMapper},label=lst:PTMapper]
java -cp pt2matsim-19.2-shaded.jar
org.matsim.pt2matsim.run.PublicTransitMapper
"path to the default configuration file" 
\end{lstlisting}
\ \\After this process has been successfully executed, all necessary data for the simulation of public transport has been created. The PT2MATSim package also offers the possibility to check implausibilities in the transit routes. Thereby it is checked whether the route has loops and whether the travel times specified in the \textit{TransitSchedule} can be achieved on the network on which the routes are mapped. The plausibility check for the schedule can be executed by the code specified in listing \ref{lst:Plausi}. The plausibility check is called by the Java main method \textit{CheckMappedSchedulePlausibility}. Subsequently, the schedule and the network files are determined. Then the coordinate system and finally the output folder are specified.
\ \\\begin{lstlisting}[language=xml, basicstyle=\footnotesize\sffamily, numbers=left, escapechar=\%, numberstyle=\tiny, frame=tb, tabsize=4, lineskip=-0.1ex, caption={Schedule plausibility check},label=lst:Plausi]
java -cp pt2matsim-19.2-shaded.jar
org.matsim.pt2matsim.run.CheckMappedSchedulePlausibility
"path to transit schedule"
"path to network file" "EPSG:code"
"path to output folder"
\end{lstlisting}
\ \\Finally, the modeling of transport demand is necessary. The population data set may include a wide range of socio-economic characteristics such as place of residence, income, place of work, among others. Depending on the availability and resolution of the data, this module differs significantly from one scenario to another \citep{kickhofer2009methodik}. As already mentioned, only a small part of the population was simulated. Socio-economic characteristics, except for the place of residence and work, were initially omitted, as these are only additional details that make the model more realistic but are not relevant for the technical implementation. Since about 1200 employees work at the Max-Planck-Institut at the Fassberg and 14.1 percent of the population live in the city centre of Göttingen, 169 pseudo agents are simulated who commute from the city centre to the Fassberg. These agents either use the car or public transport to get to work in the baseline situation. Other activities, such as shopping, were not simulated for the exemplary route. In general, however, the implementation of trip chains with MATSim is possible. Therefore, the population data for this commuting scenario to the Fassberg are fairly simple but sufficient to realize the objectives of this paper. A subset of a population file is shown in listing \ref{lst:population}. Like the network file, the population file needs to have a xml data format to be compatible with MATSim. 
\newpage
\begin{lstlisting}[language=xml, basicstyle=\footnotesize\sffamily, numbers=left, escapechar=\%, numberstyle=\tiny, frame=tb, tabsize=4, lineskip=-0.1ex, caption={Subset of a population file },label=lst:population]
<person id="pers_0">
	<plan selected="yes">
		<activity type="h" x="3565270.0" 
		y="5711985.0" end_time="08:02:45" >
		</activity>
		<leg mode="car">
		</leg>
		<activity type="w" x="3567092.2910817387" 
		y="5714427.227577877" end_time="17:22:48" >
		</activity>
		<leg mode="car">
		</leg>
		<activity type="h" x="3565270.0" y="5711985.0" >
		</activity>
	</plan>
</person>
	
<person id="pers_101">
	<plan selected="yes">
		<activity type="h" x="3565219.0" 
		y="5711797.0" end_time="08:43:50" >
		</activity>
		<leg mode="pt">
		</leg>
		<activity type="w" x="3567092.2910817387" 
		y="5714427.227577877" end_time="17:24:05" >
		</activity>
		<leg mode="pt">
		</leg>
		<activity type="h" x="3565219.0" y="5711797.0" >
		</activity>
	</plan>
</person>	
\end{lstlisting}

\ \\In order to validate the baseline scenario, the travel costs using the car and public transport should be varied. If the demand or the modal split changes according to the economic theory\footnote{According to economic theory, the demand for a normal good increases when the price decreases. In addition, the majority of Europeans agree that the best way to improve urban transport is to reduce the price of public transport \citep{AttitudesUrbanMobility}.}, it can be concluded that the baseline situation is plausible. Within the configuration file, MATSim offers several possibilities to adjust the scoring of the different means of transport. These include, for example, the marginal utility of traveling, the marginal utility of distance, and the monetary distance rate. Since the monetary distance rate describes the monetary costs per distance and therefore reflects the changes in the prices of the modes of transport, it was varied to test the baseline scenario. The results are discussed in Section \ref{sec:Results}.

\subsection{DRT Scenario}\label{sec:DRT Scenario}

The above explained baseline situation will be compared with a scenario in which there is an additional offer of DRT in the Göttingen area. This comparison is intended to build the foundation for conducting a cost-benefit analysis and assesses the choice of means of transport and its impacts on travel times, internal and external costs such as congestion, noise, and air pollution. It will then be investigated whether negative external effects can be reduced by the offer of a DRT service as a traffic measure. Achieving these long-term objectives requires establishing the technical foundations for the implementation of a DRT system in the MATSim software. The simulation of dynamic transportation modes like taxis or DRT is modeled by MATSim's extension for Dynamic Vehicle Routing Problems (DVRP), which enables the simulation of mobility on-demand. This allows a dynamic disposition of vehicle fleets during simulation in response to incoming requests and other events that differ in their objectives and constraints using pluggable algorithms \citep{bischoff2017city}. Potential restrictions include, for example, the maximum waiting time or the maximum travel time. Currently, two modes are available for simulating DRT services. Either a door-to-door service or a stop-based service. To implement the DVRP extension in MATSim, some dependencies must be added to the project object model\footnote{The project object model, named \textit{pom.xml} in MATSim, stores information for a software project supported by Maven. This file contains all the information about the software project.}. These dependencies are presented in listing \ref{lst:dependencies}. 
\ \\\begin{lstlisting}[language=xml, basicstyle=\footnotesize\sffamily, numbers=left, escapechar=\%, numberstyle=\tiny, frame=tb, tabsize=4, lineskip=-0.1ex, caption={Dependencies for the implementation of DRT},label=lst:dependencies]
<dependency>
	<groupId>org.matsim.contrib</groupId>
	<artifactId>otfvis</artifactId>
	<version>0.11.0-2018w44</version>
</dependency>

<dependency>
	<groupId>org.matsim.contrib</groupId>
	<artifactId>taxi</artifactId>
	<version>0.11.0-2018w44</version>
</dependency>

<dependency>
	<groupId>org.matsim.contrib</groupId>
	<artifactId>av</artifactId>
	<version>0.11.0-2018w44</version>
</dependency>
<dependency>
	<groupId>org.matsim.contrib</groupId>
	<artifactId>drt</artifactId>
	<version>0.11.0-2018w44</version>
</dependency>

<dependency>
	<groupId>org.matsim.contrib</groupId>
	<artifactId>dvrp</artifactId>
	<version>0.11.0-2018w44</version>
</dependency>
\end{lstlisting}

\ \\After implementation, some adjustments must be made in the configuration file, then the agents can use DRT as an additional transport mode. Those adjustments can be seen in listing \ref{lst:DRTconfig}.
\newpage
\ \\\begin{lstlisting}[language=xml, basicstyle=\footnotesize\sffamily, numbers=left, escapechar=\%, numberstyle=\tiny, frame=tb, tabsize=4, lineskip=-0.1ex, caption={Adjustments for DRT in the configuration file},label=lst:DRTconfig]
<module name="drt">
<param name="operationalScheme" value="door2door" />
<param name="stopDuration" value="60.0" />
<param name="vehiclesFile" value="path to the vehicle file " />
\end{lstlisting}
\ \\As becomes apparent from the code, the implementation of a DRT service also requires a \textit{drtvehicles} file in which the vehicle fleet is described. In this file, the departure locations of the vehicles, the operating time\footnote{In MATSim time is measured in seconds after midnight.} and the capacity are defined. In the framework of this project work, 5 DRT vehicles with 8 seats were implemented, departing from the university, the main station, and the new town hall and offering a door-to-door service. The deployed \textit{drtvehicles} file, which also needs to have a xml data format, is shown in listing \ref{lst:drtvehicle}. 
\ \\\begin{lstlisting}[language=xml, basicstyle=\footnotesize\sffamily, numbers=left, escapechar=\%, numberstyle=\tiny, frame=tb, tabsize=4, lineskip=-0.1ex, caption={drtvehicle.xml},label=lst:drtvehicle]
<vehicles>
	<vehicle id="drt0" start_link="34945-24004-24005-24006-24007-34944-31821-31822-3182" t_0="0.0" 
	t_1="86400.0" capacity="8"/>
	
	<vehicle id="drt1" start_link="2790-23646-23647-23648" t_0="0.0"
	t_1="86400.0" capacity="8"/>
	
	<vehicle id="drt2" start_link="35051-35052-35100" t_0="0.0"
	t_1="86400.0" capacity="8"/>
	
	<vehicle id="drt3" start_link="35044-35045-35046-34990-34991-34992-34993-34994-3499" t_0="0.0" 
	t_1="86400.0" capacity="8"/>
	
	<vehicle id="drt4" start_link="14899-14901-14903-5087-5089" t_0="0.0" 
	t_1="86400.0" capacity="8"/>
</vehicles>
\end{lstlisting}
\ \\Apart from the implementation of the DRT service, the requirements for the DRT scenario remain the same as for the baseline situation. However, different demands were generated for this scenario to compare DRT usage for different market shares. The simulation of a new launch of the service is compared with a situation in which the modal split of DRT usage is already 30 percent. Hence it will be derived on which means of transport the agents switch when introducing a DRT service and whether the agents make other decisions regarding the means of transport if there already exists a DRT service. These and other relevant results are discussed in the following section. 

\newpage
\section{Results}\label{sec:Results}

\begin{figure}[h!]
\centering
\includegraphics[width=0.9\linewidth]{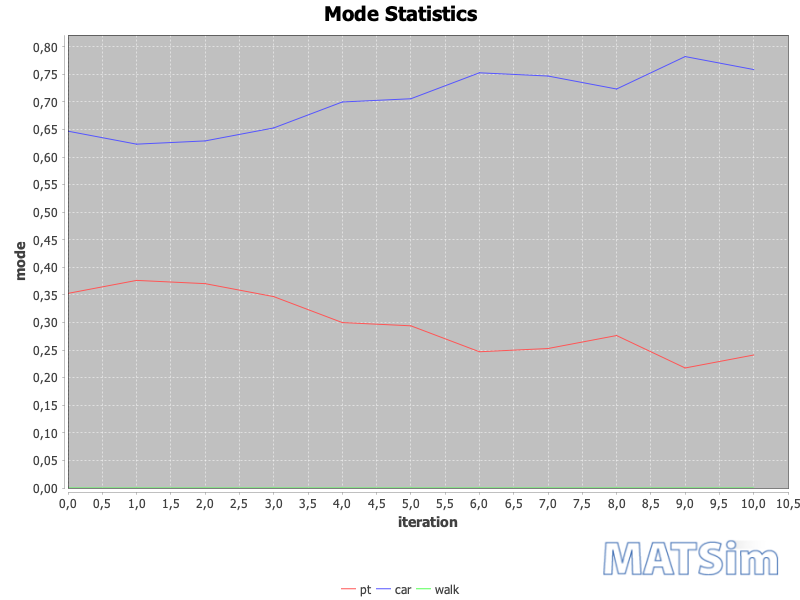}
\caption{Initial mode statistic}
\label{figure:3}
\end{figure}

This section examines the results of the simulation of the exemplary commuter route. Thereby, transport mode statistics, the variation of the monetary distance rate and the different initial demands of the DRT scenario are discussed.   
\ \\First of all, it is important to consider the preliminary transport mode statistics of the baseline situation. The transport demand was generated in such a way that at the beginning of the first iteration that simulates the baseline scenario, 65 percent of the transport demand is accounted for by car. Accordingly, the modal split for public transport is 35 percent. Figure \ref{figure:3} illustrates the initial modal split as well as the adjustments of the agents during the 10 iterations of the simulation. In the baseline scenario, more than 10 percent of agents switch from public transport to car usage. It is important to note that in this illustrative example all agents have access to all means of transport and can, therefore, change to using a car at any time without barriers. Furthermore, in this illustrative simulation, only 169 agents are on the road network, which means that the probability of traffic jams is very low. Thus, the scores of the plans where the car is used are much higher because there are no waiting times when using the car. These changes in transport mode statistics are therefore plausible.

\ \\As already described in section \ref{sec:Baseline Situation}, the monetary distance rate was varied to examine the impact of price changes on the choice of means of transport in the baseline situation. This was achieved by running four simulations using different monetary distance rates, which are illustrated in table \ref{table:1}. In the first run, no monetary distance rates are implemented for public transport as well as for the car. The corresponding transport mode statistic is plotted in figure \ref{figure:3}. In the second and third runs, the price for the distance driven by car was increased. Primarily, these runs served to get a first impression of the resulting reaction patterns. Although the monetary distance rate for the car in the third run increases a hundredfold compared to the second run, the modal split only changes slightly. This suggests that the main determinant is the relationship between the monetary distance rates of different means of transport. Consequently, in the fourth run, the monetary distance rate of the car is ten times higher than for public transport. Overall, these changes in prices have altered the choice of the modes of transport as predicted by economic theory. The baseline situation can therefore generally be regarded as plausible. It is also interesting to note that the scores of the plans for the two modes of transport are beginning to converge due to changes in the monetary distance. 

\begin{table}[h!]
\resizebox{1.12\textwidth}{!}{\begin{minipage}{\textwidth}
\begin{tabular} {|c|c|c|l|l|l|l|c|c|}
\hline
iteration 
& \multicolumn{8}{c|}{run} \\ \hline
& \multicolumn{2}{c|}{1} 
& \multicolumn{2}{c|}{\cellcolor[HTML]{EFEFEF}2} 
& \multicolumn{2}{c|}{3} 
& \multicolumn{2}{c|}{\cellcolor[HTML]{EFEFEF}4} \\ \hline
\multicolumn{1}{|l|}{\begin{tabular}[c]{@{}l@{}}monetary \\ distance \\ rate:\end{tabular}} 
& \begin{tabular}[c]{@{}c@{}}car\\ \\ \\ \\ 0\end{tabular} 
& \begin{tabular}[c]{@{}c@{}}pt\\ \\ \\ \\ 0\end{tabular} 
& \multicolumn{1}{c|}{\cellcolor[HTML]{EFEFEF}\begin{tabular}[c]{@{}c@{}}car\\ \\ \\ \\ -0.2\end{tabular}} 
& \multicolumn{1}{c|}{\cellcolor[HTML]{EFEFEF}\begin{tabular}[c]{@{}c@{}}pt\\ \\ \\ \\ 0\end{tabular}} 
& \multicolumn{1}{c|}{\begin{tabular}[c]{@{}c@{}}car\\ \\ \\ \\ -0.002\end{tabular}} 
& \multicolumn{1}{c|}{\begin{tabular}[c]{@{}c@{}}pt\\ \\ \\ \\ 0\end{tabular}} 
& \cellcolor[HTML]{EFEFEF}\begin{tabular}[c]{@{}c@{}}car\\ \\ \\ \\ -0.00154\end{tabular} 
& \cellcolor[HTML]{EFEFEF}\begin{tabular}[c]{@{}c@{}}pt\\ \\ \\ \\ -0.000154\end{tabular} \\ \hline
0 
& 65 \% 
& 35 \% 
& \multicolumn{1}{c|}{\cellcolor[HTML]{EFEFEF}65 \%} 
& \cellcolor[HTML]{EFEFEF}35 \% 
& \multicolumn{1}{c|}{65 \%} 
& \multicolumn{1}{c|}{35 \%} 
& \cellcolor[HTML]{EFEFEF}65 \% 
& \cellcolor[HTML]{EFEFEF}35 \% \\ \hline
1 
& 62 \% 
& 38 \% 
& \multicolumn{1}{c|}{\cellcolor[HTML]{EFEFEF}62 \%} 
& \cellcolor[HTML]{EFEFEF}38 \% 
& 62 \% 
& 38 \% 
& \cellcolor[HTML]{EFEFEF}62 \% 
& \cellcolor[HTML]{EFEFEF}38 \% \\ \hline
2 
& 63 \% 
& 37 \% 
& \cellcolor[HTML]{EFEFEF}58 \% 
& \cellcolor[HTML]{EFEFEF}42 \% 
& 58 \% 
& 42 \% 
& \cellcolor[HTML]{EFEFEF}59 \% 
& \cellcolor[HTML]{EFEFEF}41 \% \\ \hline
3 
& 65 \% 
& 35 \% 
& \cellcolor[HTML]{EFEFEF}52 \% 
& \cellcolor[HTML]{EFEFEF}48 \% 
& 54 \% 
& 46 \% 
& \cellcolor[HTML]{EFEFEF}57 \% 
& \cellcolor[HTML]{EFEFEF}43 \% \\ \hline
4 
& 70 \% 
& 30 \% 
& \cellcolor[HTML]{EFEFEF}49 \% 
& \cellcolor[HTML]{EFEFEF}51 \% 
& 51 \% 
& 49 \% 
& \cellcolor[HTML]{EFEFEF}55 \% 
& \cellcolor[HTML]{EFEFEF}45 \% \\ \hline
5 
& 71 \% 
& 29 \% 
& \cellcolor[HTML]{EFEFEF}45 \% 
& \cellcolor[HTML]{EFEFEF}55 \% 
& 47 \% 
& 53 \% 
& \cellcolor[HTML]{EFEFEF}51 \% 
& \cellcolor[HTML]{EFEFEF}49 \% \\ \hline
6 
& \multicolumn{1}{l|}{75 \%} 
& \multicolumn{1}{l|}{25 \%} 
& \cellcolor[HTML]{EFEFEF}42 \% 
& \cellcolor[HTML]{EFEFEF}58 \% 
& 44 \%
& 56 \% 
& \cellcolor[HTML]{EFEFEF}49 \% 
& \cellcolor[HTML]{EFEFEF}51 \% \\ \hline
7 
& \multicolumn{1}{l|}{75 \%} 
& \multicolumn{1}{l|}{25 \%} 
& \cellcolor[HTML]{EFEFEF}39 \% 
& \cellcolor[HTML]{EFEFEF}61 \% 
& 42 \% 
& 58 \% 
& \cellcolor[HTML]{EFEFEF}48 \% 
& \cellcolor[HTML]{EFEFEF}52 \% \\ \hline
8 
& \multicolumn{1}{l|}{72 \%} 
& \multicolumn{1}{l|}{28 \%} 
& \cellcolor[HTML]{EFEFEF}38 \% 
& \cellcolor[HTML]{EFEFEF}62 \% 
& 42 \% 
& 58 \% 
& \cellcolor[HTML]{EFEFEF}48 \% 
& \cellcolor[HTML]{EFEFEF}52 \% \\ \hline
9 
& \multicolumn{1}{l|}{78 \%} 
& \multicolumn{1}{l|}{22 \%} 
& \cellcolor[HTML]{EFEFEF}38 \% 
& \cellcolor[HTML]{EFEFEF}62 \% 
& 42 \% 
& 58 \% 
& \cellcolor[HTML]{EFEFEF}48 \% 
& \cellcolor[HTML]{EFEFEF}52 \% \\ \hline
10 
& \multicolumn{1}{l|}{76 \%} 
& \multicolumn{1}{l|}{24 \%} 
& \cellcolor[HTML]{EFEFEF}34 \% 
& \cellcolor[HTML]{EFEFEF}66 \% 
& 36 \% 
& 64 \% 
& \cellcolor[HTML]{EFEFEF}45 \% 
& \cellcolor[HTML]{EFEFEF}55 \% \\ \hline
\end{tabular}
\end{minipage} }
\caption{Simulation runs with different monetary distance rates}
\label{table:1}
\end{table}

\ \\Finally, the results of the DRT scenario simulation should be
considered more closely. As previously mentioned, two different demands
were generated for the scenario to compare the usage of DRT for
different market shares. The results are summarized in table
\ref{table:2}. First of all, it needs to be pointed out that, in the
situation of a new introduction of a DRT service, the users of the
public transport system are primarily switching to the door-to-door
offer. Compared to the situation without the provision of DRT, the
share of public transport in the modal split is even reduced by two
percent. Moreover, in contrast to the situation without DRT, car
traffic does not increase over the ten iterations of the simulation.
Due to the introduction of the new DRT service there are altogether
fifteen cars less on the network\footnote{For comparison, run 1 of the baseline situation 
is considered, since in the DRT scenario the
monetary distance rate for all means of transport is set to zero.}. Similarly,
in a situation where DRT already has a 30 percent share of the modal
split, there are fewer cars on the network. It is interesting to note that this simulation 
reveals the smallest fluctuations in the share of DRT and that public transport users are
more likely to switch to cars than to DRT. The scores obtained for the tours covered with
DRT are very close to the scores for the tours covered by cars. However, the majority of
agents decide for car usage, as it often performs slightly better. It can, therefore, be
assumed that the mode choice will be different if the
appropriate costs are properly integrated into the software. This significant 
implementation and other important next steps for a realistic traffic simulation are 
explained in the next section. 

\begin{table}[h!]
\resizebox{1\textwidth}{!}{\begin{minipage}{\textwidth}
\centering
\begin{tabular}{|c|c|c|c|c|c|c|}
\hline
iteration & \multicolumn{3}{c|}{Initially no DRT} & \multicolumn{3}{c|}{\begin{tabular}[c]{@{}c@{}}Initial share \\ of about 30 \% DRT\end{tabular}} \\ \hline
 & car & DRT & pt & car & DRT & pt \\ \hline
\rowcolor[HTML]{EFEFEF} 
0 & 65 \% & 0 \% & 35 \% & 45 \% & 29 \% & 26 \% \\ \hline
1 & 61 \% & 4 \% & 35 \% & 42 \% & 31 \% & 28 \% \\ \hline
\rowcolor[HTML]{EFEFEF} 
2 & 59 \% & 9 \% & 32 \% & 43 \% & 29 \% & 28 \% \\ \hline
3 & 61 \% & 9 \% & 31 \% & 44 \% & 32 \% & 24 \% \\ \hline
\rowcolor[HTML]{EFEFEF} 
4 & 63 \% & 9 \% & 28 \% & 51 \% & 27 \% & 22 \% \\ \hline
5 & 62 \% & 13 \% & 25 \% & 48 \% & 32 \% & 19 \% \\ \hline
\rowcolor[HTML]{EFEFEF} 
6 & 67 \% & 12 \% & 21 \% & 56 \% & 27 \% & 17 \% \\ \hline
7 & 65 \% & 15 \% & 20 \% & 56 \% & 27 \% & 17 \% \\ \hline
\rowcolor[HTML]{EFEFEF} 
8 & 64 \% & 14 \% & 22 \% & 54 \% & 28 \% & 18 \% \\ \hline
9 & 68 \% & 15 \% & 17 \% & 54 \% & 30 \% & 16 \% \\ \hline
\rowcolor[HTML]{EFEFEF} 
10 & 64 \% & 14 \% & 22 \% & 59 \% & 26 \% & 15 \% \\ \hline
\end{tabular}
\caption{Simulation runs with different initial shares of DRT usage}
\label{table:2}
\end{minipage} }
\end{table}

\section{Next steps}

This section describes the necessary extensions and the next steps for comprehensive traffic simulation and a subsequent cost-benefit analysis of a DRT offer in Göttingen and the surrounding area. Within the scope of a next paper, multi-agent simulations based on real survey data from Göttingen, Bovenden and Rosdorf will be carried out. Therefore, it becomes necessary to generate agent activity plans from mobility data. There are different approaches for generating a synthetic agent population with socio-economic data for each agent by integrating different data sources. Related work aimed at generating a close-to-reality population is often based on iterative proportional fitting designs with additional Monte Carlo sampling. Examples of this are studies by \citet{muller2011hierarchical} and \citet{zhu2014synthetic}. A further approach regarding the generation of synthetic population is the application of the concept of discrete choice models as in \citet{anderson2014associations}. Besides the generation of a synthetic agent population, the implementation of the bicycle as a transportation mode is very important to derive a realistic traffic forecast, as the current city-wide modal share of cycling ranges at 28 \% in Göttingen \citep{MobilitaetGoettingen}. Therefore, it is necessary to extend the MATSim framework with cycling as a means of transport. A proposal for implementation is made by \citet{ziemke2017modeling}. In addition, for a cost-benefit analysis, it is necessary to integrate some extensions of the MATSim software. An essential feature of the analysis is the monetization of negative external effects such as emissions. Therefore, it is necessary, for example, to implement the emission package for MATSim. Thereby emissions are calculated every time the vehicle leaves the road and costs vary for different engine types, ages, displacement, and traffic conditions, e.g. free flow or congestion \citep{MATSimEmissions}. In addition, the time required by motorists to find a parking lot should be incorporated, as otherwise the costs of driving would not be fully reflected \citep{horni2013agent}. 
Other factors relevant for the cost-benefit analysis, such as travel time, waiting time, and internal costs such as fuel or ticket prices for the various means of transport, can be integrated into the scoring function as already indicated in this project work. For a meaningful calibration of the scoring function parameters of a mode choice model can be used. Finally, it would be helpful to extend the software with some analytical tools. The only statistic that comes out of the box is the mode statistic, but a comparison of the scores, the travel times and the traveled  distances for the different means of transport would be very helpful and important for a comprehensive analysis of the topic. A comprehensive cost-benefit analysis of a DRT offer should be possible by adding these crucial next steps to the simulation scenarios. Accordingly, the conclusion can be drawn. 

\section{Conclusion}

This paper investigates the requirements of the technical implementation of cost-benefit analysis with the transport simulation software MATSim. As an innovative tool, such simulations could effectively support the work of planning authorities in the design of transport projects in the future. For example, it could be investigated what proportion of the population would support the project under consideration. To test the method of traffic simulation a small illustrative scenario was designed before applying it to a real world scenario in an upcoming study. It was shown how to set up a basic simulation and how this can be extended with public transport and DRT. Besides, it was demonstrated how the scenarios could be varied, for example by the monetary distance rates of different means of transport. Finally, the results were presented and the necessary next steps were explained. The presented approach should be seen as a demonstration of the capabilities of the multi-agent transport simulation software MATSim. The possibilities of the use are manifold and seem to be arbitrarily expandable. Altogether it can be concluded that the software is suitable for a cost-benefit analysis of a DRT service. Above all, the most important internal and external costs, such as usage costs of the various modes of transport and emissions, can be integrated into the simulation scenarios, enabling an evaluation of the costs and benefits of a DRT service when using mobile data from Göttingen and the surrounding area. 

\section*{Acknowledgements}
We gratefully acknowledge the funding from the European Union’s Horizon 2020 research and innovation programme under the Marie Skłodowska-Curie grant agreement No 800267.

\cleardoublepage
\pagenumbering{Roman}
\setcounter {page} {5}
\addcontentsline {toc} {section} {References}
\bibliography{lit}

\begin{thebibliography}{}

\bibitem[Agarwal et~al., 2019]{agarwal2019calibration}
Agarwal, A., Ziemke, D., and Nagel, K. (2019).
\newblock Calibration of choice model parameters in a transport scenario with
  heterogeneous traffic conditions and income dependency.
\newblock {\em Transportation Letters}, pages 1--10.

\bibitem[Anderson et~al., 2014]{anderson2014associations}
Anderson, P., Farooq, B., Efthymiou, D., and Bierlaire, M. (2014).
\newblock Associations generation in synthetic population for transportation
  applications: Graph-theoretic solution.
\newblock {\em Transportation Research Record}, 2429(1):38--50.

\bibitem[Bischoff and Maciejewski, 2016]{bischoff2016simulation}
Bischoff, J. and Maciejewski, M. (2016).
\newblock Simulation of city-wide replacement of private cars with autonomous
  taxis in berlin.
\newblock {\em Procedia computer science}, 83:237--244.

\bibitem[Bischoff et~al., 2017]{bischoff2017city}
Bischoff, J., Maciejewski, M., and Nagel, K. (2017).
\newblock City-wide shared taxis: A simulation study in berlin.
\newblock In {\em 2017 IEEE 20th International Conference on Intelligent
  Transportation Systems (ITSC)}, pages 275--280. IEEE.

\bibitem[Bristow and Nellthorp, 2000]{bristow2000transport}
Bristow, A. and Nellthorp, J. (2000).
\newblock {Transport project appraisal in the European Union}.
\newblock {\em Transport Policy}, 7(1):51--60.

\bibitem[Ciari et~al., 2008]{ciari2008concepts}
Ciari, F., Balmer, M., and Axhausen, K.~W. (2008).
\newblock Concepts for a large scale car-sharing system: Modelling and
  evaluation with an agent-based approach.
\newblock {\em Working paper/IVT}, 517.

\bibitem[Cich et~al., 2017]{cich2017modeling}
Cich, G., Knapen, L., Maciejewski, M., Yasar, A.-U.-H., Bellemans, T., and
  Janssens, D. (2017).
\newblock Modeling demand responsive transport using sarl and matsim.
\newblock Elsevier BV.

\bibitem[Damart and Roy, 2009]{damart2009uses}
Damart, S. and Roy, B. (2009).
\newblock {The uses of cost--benefit analysis in public transportation
  decision-making in France}.
\newblock {\em Transport Policy}, 16(4):200--212.

\bibitem[G{\"o}ssling and Choi, 2015]{gossling2015transport}
G{\"o}ssling, S. and Choi, A.~S. (2015).
\newblock {Transport transitions in Copenhagen: Comparing the cost of cars and
  bicycles}.
\newblock {\em Ecological Economics}, 113:106--113.

\bibitem[Horni et~al., 2013]{horni2013agent}
Horni, A., Montini, L., Waraich, R.~A., and Axhausen, K.~W. (2013).
\newblock An agent-based cellular automaton cruising-for-parking simulation.
\newblock {\em Transportation Letters}, 5(4):167--175.

\bibitem[Horni et~al., 2016]{horni2016multi}
Horni, A., Nagel, K., and Axhausen, K.~W. (2016).
\newblock {\em The multi-agent transport simulation MATSim}.
\newblock Ubiquity Press London.

\bibitem[Johansson et~al., 2017]{johansson2017evaluation}
Johansson, E., Winslott~Hiselius, L., Koglin, T., and Wretstrand, A. (2017).
\newblock {Evaluation of public transport: regional policies and planning
  practices in Sweden}.
\newblock {\em Urban, Planning and Transport Research}, 5(1):59--77.

\bibitem[Kaddoura et~al., 2015]{kaddoura2015optimal}
Kaddoura, I., Kickh{\"o}fer, B., Neumann, A., and Tirachini, A. (2015).
\newblock Optimal public transport pricing: Towards an agent-based marginal
  social cost approach.
\newblock {\em Journal of Transport Economics and Policy (JTEP)},
  49(2):200--218.

\bibitem[Kickh{\"o}fer, 2009]{kickhofer2009methodik}
Kickh{\"o}fer, B. (2009).
\newblock {Die Methodik der {\"o}konomischen Bewertung von
  Verkehrsma{\ss}nahmen in Multiagentensimulationen}.

\bibitem[{LK Argus}, 2016]{MobilitaetGoettingen}
{LK Argus} (2016).
\newblock {Haushaltsbefragung zum Mobilitätsverhlaten 2015 -
  Göttingen/Rosdorf/Bovenden}.
\newblock Technical report, {Stadt Göttingen and Zweckverband Verkehrsverbund
  Süd-Nidersachsen}.

\bibitem[Mackie and Worsley, 2013]{mackie2013international}
Mackie, P. and Worsley, T. (2013).
\newblock International comparisons of transport appraisal practice: Overview
  report.
\newblock {\em Institute for Transport Studies, University of Leeds, Leeds}.

\bibitem[MATSim, 2019]{MATSimEmissions}
MATSim (2019).
\newblock \emph{{MATSim - Multi-Agent Transport Simulation - Munich, Germany}}.
\newblock \url{https://www.matsim.org/gallery/munich} [Accessed: 18.07.2019].

\bibitem[M{\"u}ller and Axhausen, 2011]{muller2011hierarchical}
M{\"u}ller, K. and Axhausen, K.~W. (2011).
\newblock Hierarchical ipf: Generating a synthetic population for switzerland.
\newblock {\em Arbeitsberichte Verkehrs-und Raumplanung}, 718.

\bibitem[Nagel and Fl{\"o}tter{\"o}d, 2012]{nagel2012agent}
Nagel, K. and Fl{\"o}tter{\"o}d, G. (2012).
\newblock Agent-based traffic assignment: Going from trips to behavioural
  travelers.
\newblock In {\em Travel Behaviour Research in an Evolving World--Selected
  papers from the 12th international conference on travel behaviour research},
  pages 261--294. International Association for Travel Behaviour Research.

\bibitem[Poletti, 2016]{poletti2016public}
Poletti, F. (2016).
\newblock {Public Transit Mapping on Multi-Modal Networks in MATSim}.
\newblock Master's thesis, ETH Zurich.

\bibitem[Rieser, 2010]{rieser2010adding}
Rieser, M. (2010).
\newblock Adding transit to an agent-based transportation simulation.
\newblock {\em Swiss Federal Institute of Technology, Z{\"u}rich}.

\bibitem[{TNS Opinion and DG MOVE}, 2013]{AttitudesUrbanMobility}
{TNS Opinion and DG MOVE} (2013).
\newblock {Special Eurobarometer 406 - Attitudes of Europeans towards urban
  mobility}.
\newblock Technical report, {European Commission}.

\bibitem[UBA, 2019]{UBAFahrleistung}
UBA (2019).
\newblock \emph{{Fahrleistungen, Verkehrsaufwand und „Modal Split“}}.
\newblock
  \url{https://www.umweltbundesamt.de/daten/verkehr/fahrleistungen-verkehrsaufwand-modal-split#textpart-1}
  [Accessed: 06.06.2019].

\bibitem[Zhu and Ferreira, 2014]{zhu2014synthetic}
Zhu, Y. and Ferreira, J. (2014).
\newblock Synthetic population generation at disaggregated spatial scales for
  land use and transportation microsimulation.
\newblock {\em Transportation Research Record}, 2429(1):168--177.

\bibitem[Ziemke et~al., 2017]{ziemke2017modeling}
Ziemke, D., Metzler, S., and Nagel, K. (2017).
\newblock Modeling bicycle traffic in an agent-based transport simulation.
\newblock {\em Procedia Computer Science}, 109:923--928.

\end{thebibliography}

\end{document}